\documentclass{elsart}

 \usepackage{graphicx}

\usepackage{amssymb}

\begin{document}

\begin{frontmatter}



\title{\boldmath Self energies of the pion and the 
$\Delta$ isobar from the  
 $^{3}$He(e,e'$\pi^{+}$)$^3$H reaction}


\author[Darmstadt]{M.~Kohl},
\author[Mainz]{P.~Bartsch},
\author[Mainz]{D.~Baumann},
\author[MainzP]{J.~Bermuth},
\author[Mainz]{R.~B\"{o}hm},
\author[Ljubljana]{K.~Bohinc},
\author[Mainz]{S.~Derber},
\author[Mainz]{M.~Ding},
\author[Mainz]{M.O.~Distler},
\author[Mainz]{I.~Ewald},
\author[Mainz]{J.~Friedrich},
\author[Mainz]{J.M.~Friedrich},
\author[Mainz]{P.~Jennewein},
\author[Mainz]{M.~Kahrau},
\author[Dubna]{S.S.~Kamalov},
\author[Melbourne]{A.~Kozlov},
\author[Mainz]{K.W.~Krygier},
\author[Pisa]{M.~Kuss},
\author[Mainz]{A.~Liesenfeld},
\author[Mainz]{H.~Merkel},
\author[Mainz]{P.~Merle},
\author[Mainz]{U.~M\"{u}ller},
\author[Mainz]{R.~Neuhausen},
\author[Mainz]{Th.~Pospischil},
\author[Ljubljana]{M.~Potokar},
\author[Saskatoon]{C.~Rangacharyulu},
\author[Darmstadt]{A.~Richter\corauthref{cor1}},
\corauth[cor1]{Corresponding author.}
\ead{richter@ikp.tu-darmstadt.de}
\author[Basel]{D.~Rohe},
\author[Glasgow]{G.~Rosner},
\author[Mainz]{H.~Schmieden},
\author[Mainz]{G.~Schrieder},
\author[Mainz]{M.~Seimetz},
\author[MIT]{S.~\v{S}irca},
\author[Riken]{T.~Suda},
\author[Mainz]{L.~Tiator},
\author[Darmstadt]{M.~Urban},
\author[Mainz]{A.~Wagner},
\author[Mainz]{Th.~Walcher},
\author[Darmstadt]{J.~Wambach},
\author[Mainz]{M.~Weis},
\author[Juelich]{A.~Wirzba}

\address[Darmstadt]{Institut f\"{u}r Kernphysik, 
    Technische Universit\"at Darmstadt, D-64289 Darmstadt,
    Germany}
\address[Mainz]{Institut f\"{u}r Kernphysik, 
    Universit\"{a}t Mainz,  D-55099 Mainz,
    Germany}
\address[MainzP]{Institut f\"{u}r Physik, 
    Universit\"{a}t Mainz,  D-55099 Mainz,
    Germany}
\address[Ljubljana]{Institute ``Jo\v{z}ef Stefan'', 
    University of Ljubljana,
    SI-1001 Ljubljana, Slovenia}
\address[Dubna]{Laboratory of Theoretical Physics, JINR Dubna,
    SU-10100 Moscow, Russia}
\address[Melbourne]{School of Physics, The University of Melbourne,
    Victoria 3010, Australia}
\address[Pisa]{INFN Sezione di Pisa, 56010 San Piero a Grado, Italy}
\address[Saskatoon]{Department of Physics, 
    University of Saskatchewan, Saskatoon,
    SK, S7N 5E2, Canada}
\address[Basel]{Institut f\"ur Physik, Universt\"at Basel, CH-4056 Basel, 
    Switzerland}
\address[Glasgow]{Department of Physics and Astronomy, University of Glasgow,
    Glasgow G12 8QQ, UK}
\address[MIT]{Laboratory for Nuclear Science,
    Massachusetts Institute of Technology, Cambridge, MA, USA}
\address[Riken]{RI-Beam Science, RIKEN, 2-1, 
    Hirosawa, Wako, Saitama, 351-0198, 
    Japan}
\address[Juelich]{Institut f\"{u}r Kernphysik, 
    Forschungszentrum J\"{u}lich,
    D-52425 J\"ulich, Germany}

\begin{abstract}
In a kinematically complete experiment at
the Mainz microtron MAMI, pion angular 
distributions of the $^3$He(e,e'$\pi^+$)$^3$H
reaction have been measured in the excitation region
of the $\Delta$ resonance to determine the
longitudinal ($L$), transverse ($T$), and the $LT$
interference part of the differential cross section.
The data are described only after
introducing self-energy modifications of the pion and
$\Delta$-isobar propagators. Using Chiral Perturbation
Theory (ChPT) to extrapolate the pion self energy as 
inferred from the measurement on the mass shell, we
deduce a reduction of the $\pi^+$ mass of 
$\Delta m_{\pi^+} = \left(-1.7^{\;+\;1.7}_{\;-\;2.1}\right)$ MeV/c$^2$
in the neutron-rich nuclear medium
at a density of $\rho = \left(0.057^{\;+\;0.085}_{\;-\;0.057}\right)$ 
fm$^{-3}$.
Our data are consistent with the $\Delta$ self energy determined from 
measurements of $\pi^0$ photoproduction from $^4$He and heavier
nuclei.
\end{abstract}

\begin{keyword}
Pion Electroproduction \sep Longitudinal-Transverse Separation \sep
Few-Body System \sep $^3$He \sep Medium Effects \sep Delta Resonance Region
\sep Self Energy

\PACS 21.45.+v \sep 25.10.+s \sep 25.30.Rw \sep 27.10.+h
\end{keyword}
\end{frontmatter}

\section{Introduction}
\label{sec:introduction}

A basic question in hadronic physics concerns the properties of
constituents  as  they are embedded in a nuclear medium.
Such medium effects are commonly treated in terms of 
self energies from which effective masses and decay widths are deduced.
Electroproduction of charged pions from $^3$He
represents a viable testing
ground to study the influence of the nuclear medium on the production
and propagation of mesons and nucleon resonances such as the pion and
the $\Delta$ resonance.
As a simple composite nucleus, $^3$He is amenable 
to precise microscopic calculations of the wave
function and other ground state properties~\cite{mar98} and
offers the great advantage that effects of final state 
interaction are expected to be much smaller than in heavier nuclei.
Moreover, the mass-three nucleus may already be considered as a medium.
In this letter, we present 
the results of an experiment which allows the determination
of the self energies of the pion and the $\Delta$ isobar
from the analysis of the longitudinal and transverse cross section
components, respectively.
These self-energy terms are the subject of theoretical descriptions in
the framework of the $\Delta$-hole model~\cite{delta-hole} and Chiral
Perturbation Theory (ChPT)~\cite{leut}.

\section{Measurements}
\label{sec:experiment}

To this end, we have measured the $^3$He(e,e'$\pi^+$)$^3$H reaction in 
a kinematically complete experiment at the high-resolution three-spectrometer
facility~\cite{MAMI} 
of the A1 collaboration at the 855 MeV Mainz microtron (MAMI). 
The specific experimental arrangements of the present experiment, 
including that
of the cryogenic gas target and the data acquisition and analyses  methods 
are described in detail in~\cite{blo97}. 
The very high missing
mass resolution of $\delta M \approx 700$ 
keV/c$^{2}$ (FWHM) is quite
adequate~\cite{blo97,kohl,blo96,tai00}
to clearly isolate
the coherent channel ($^3$H$\pi^+$) 
from the three- and four-body final 
states (nd$\pi^+$) and (nnp$\pi^+$).

The three-fold differential pion electroproduction cross section 
with   unpolarized electron beam
and target can be written as~\cite{kam_pi0he4}
\begin{eqnarray*}
{{d^3\sigma  }\over {d\Omega_{e'} d E_{e'} d\Omega_\pi }}=
\,\Gamma \;{{d\sigma_V }\over {d\Omega_\pi }}(W,Q^{2},\theta_{\pi};\,
\phi_{\pi}, \epsilon) 
\end{eqnarray*}
{\rm with}
\begin{eqnarray} 
{{d\sigma_V }\over {d\Omega_\pi}} & = &
{{d\sigma_T }\over {d\Omega_\pi}} + \epsilon 
{{d\sigma_L }\over {d\Omega_\pi}}
 + \sqrt{2\epsilon 
(1+\epsilon)} \,  \cos \phi_\pi\, {{d\sigma_{LT}}\over {d\Omega_\pi}}
+ \epsilon \cos 2 \phi_\pi \, { {d\sigma_{TT}}\over {d\Omega_\pi}} \, .
\label{eq:crsct}
\end{eqnarray}
Here the quantities $\epsilon $ and $\Gamma$ denote 
the polarization and flux of the virtual photon.
The indices $T$, $L$, $LT$, and
$TT$ refer to the transverse and longitudinal components and their
interferences, respectively.
The explicit dependence of $d \sigma_{V} / d
\Omega_{\pi}$ on the azimuthal pion angle $\phi_{\pi}$ and the
polarization $\epsilon$ is used for a separation of the response functions.

The measurements were carried out at two four-momentum transfers
$Q^2= 0.045$  
and 0.100 (GeV/c)$^2$, referred to as kinematics 1 and 2, respectively. 
The energy transfer in the laboratory frame has been chosen at
$\omega = 400$ and 394 MeV, respectively,
i.e. in the $\Delta$ resonance region.  At each $Q^2$,
three measurements in parallel kinematics with various values of 
$\epsilon$ were made to 
determine the $L$ and $T$
cross sections (Rosenbluth separation).  
Parallel kinematics implies that the pion is detected in the direction of the
three momentum of the virtual photon.
We have also measured the in-plane pion angular distribution
(i.e. $\phi_\pi=0^\circ$ or $180^\circ$, respectively) for the second
kinematics at $\epsilon=0.74$
to determine the $LT$ 
term. Parts of the experimental results together with
model interpretations
have already been
presented elsewhere~\cite{blo97,tai00}.  In this letter, we 
offer a combined analysis of the entire data set 
of the experiments in the two kinematics and draw definitive
conclusions about medium effects, which are especially well understood
for the pion.

\section{Results and Discussion}
\label{sec:results}
\begin{figure}[hb]
\centering\includegraphics[angle=0,width=\textwidth]{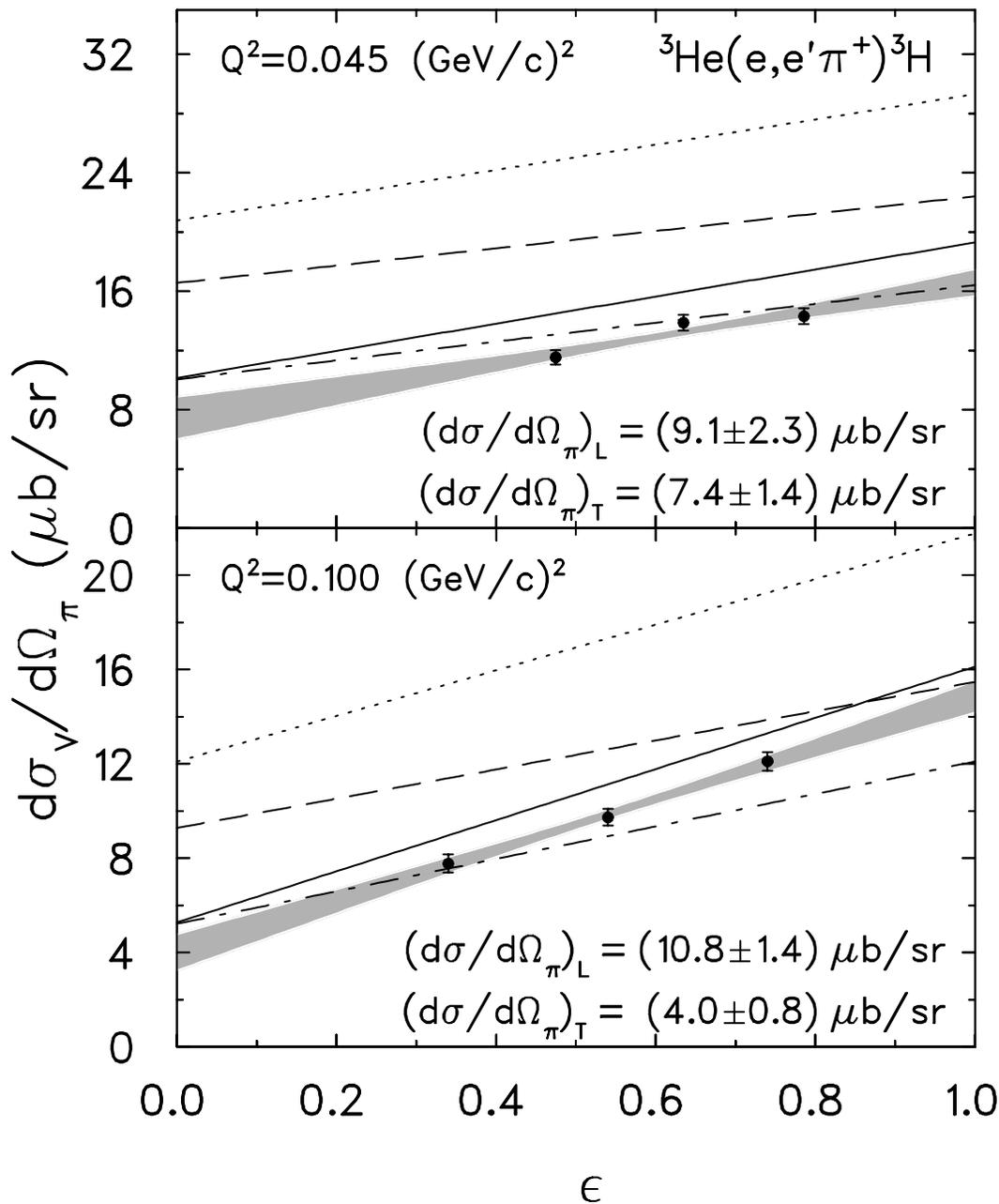}
\vspace*{3ex}
\caption{Rosenbluth plots of cross sections (Eq.~(\ref{eq:crsct})).
The data are shown as solid dots. The shaded areas are error bands
of a straight line fit to the data.
Also shown are the fit results for
the $L$ and $T$ components with statistical errors.
The dotted and dashed lines are PWIA and DWIA results, respectively. 
The dash-dotted lines include the $\Delta$ self-energy
term, while the solid lines contain 
both the $\Delta$ and pion self-energy terms.
}
\label{fig:rosen}
\end{figure}

The results of the Rosenbluth separation are shown in Fig.~\ref{fig:rosen}
where the cross sections are displayed as a function of 
the virtual photon polarization. 
The longitudinal cross section is identified as the slope,
while the transverse one is given by the
intercept with the axis at $\epsilon = 0$.  
Also shown in Fig.~\ref{fig:rosen} are the fit results for
the $L$ and $T$ components with statistical errors. The systematic
errors 
amount to 10~\% (8~\%) for
kinematics 1 (2), respectively.
The theoretical calculations
are based on the most recent elementary pion
production amplitude in the framework of the so-called Unitary Isobar 
Model~\cite{kam_pi0he4,uim99}. In Plane-Wave Impulse Approximation (PWIA),
the amplitude includes the Born terms as well as $\Delta$-
and higher resonance terms.  For the mass-three nuclei,
realistic three-body Faddeev wave functions are employed.
In the Distorted-Wave (DWIA)
calculations, the final state interaction due to pion
rescattering is included~\cite{kam93}.
As is seen in Fig.~\ref{fig:rosen},
the DWIA calculations underestimate the longitudinal component 
and overestimate the transverse component,
each by about a factor of two. Since the 
longitudinal component is dominated by the pion-pole term
and a large part of the transverse part arises from
the $\Delta$ resonance excitation,
both the pion and the $\Delta$
propagators have to be modified (see also \cite{gil97}).
In parallel kinematics the pion-pole term only contributes to the longitudinal
part of the cross section, while the $\Delta$ excitation is almost purely
transverse. 
Therefore the pion-pole and the $\Delta$ contribution
essentially decouple in the longitudinal and transverse channel and
can be studied separately.
We next discuss the estimate of these terms.

\subsection{Modification of the Pion}
\label{sec:pionmod}
\begin{figure}[b]
\centering
\vspace*{2ex}
\includegraphics[angle=90,width=\textwidth]{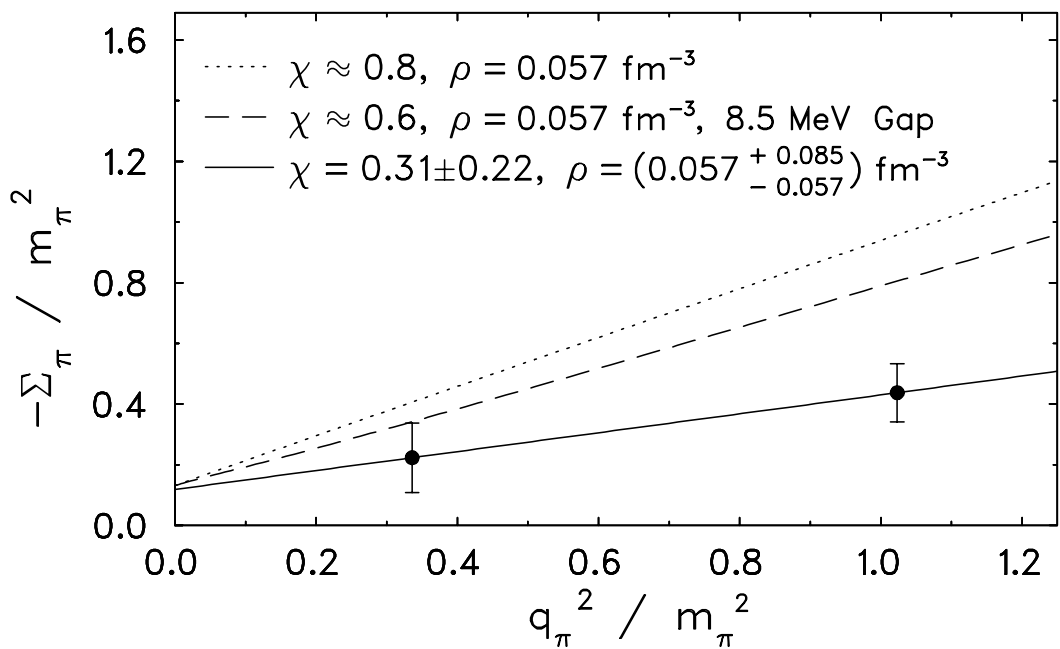}
  \caption{
The pion self energy as a function of $\vec{q}_\pi^{\;\,2}$ near
$\omega_\pi \approx 0$. The data points with the error bars
are from the longitudinal cross sections.
The dotted line corresponds to $\chi \approx 0.8$ from
the Lindhard function
with $\rho = 0.057$~fm$^{-3}$. The dashed line results after taking into
account a gap of 8.5 MeV for the ph excitation energy, 
i.e. the binding energy of $^{3}$H. The solid line
results from a fit of $\chi$ and $\rho$ to the data according to
Eq.~(\ref{eq:pionself1}).
}
  \label{fig:pi}
\end{figure}

The inadequacy of the DWIA to account for the longitudinal response 
(cf. Fig.~\ref{fig:rosen}) is remedied by replacing the free pion propagator
in the t-channel pion-pole term of the elementary amplitude,
$[\omega_\pi^2 - \vec{q\,}_\pi^2 - m_\pi^2]^{-1}$, by a modified one, 
$[\omega_\pi^2 - \vec{q\,}_\pi^2 - m_\pi^2
  - \Sigma_\pi(\omega_\pi,\vec{q}_\pi)]^{-1}$,
where $\Sigma_\pi(\omega_\pi,\vec{q}_\pi)$ denotes the pion self energy in
the nuclear medium~\cite{ErWe}. 
For the two values of $Q^2$, the
energy $\omega_\pi$ and the momentum $\vec{q}_\pi$ of the virtual pion
are fixed as $\omega_\pi =$ 1.7 (4.1) MeV and
$|\vec{q}_\pi| =$ 80.9 (141.2) MeV/c, such that two experimental numbers for 
$\Sigma_\pi$ can be determined from a fit to
the respective longitudinal cross sections. The best-fit values result in 
$\Sigma_\pi = -(0.22 \pm 0.11) \,m_\pi^2$ for kinematics 1 and
$\Sigma_\pi = -(0.44 \pm 0.10) \,m_\pi^2$ for kinematics 2.
Close to the static limit, i.e. for $\omega_\pi \approx 0$, appropriate for the
kinematical conditions of the present experiment, the pion self energy can 
be written as
\begin{equation}
\label{eq:pionself1}
\Sigma_\pi (0, \vec{q}_\pi) =
  -\frac{\sigma_N}{f_\pi^{\,2}}(\rho_p+\rho_n)
  -\vec{q}_{\pi}^{\;2} \,\chi(0,\vec{q}_\pi),
\end{equation}
where $\rho_p$ and $\rho_n$ denote the proton and neutron densities,
$\sigma_N = 45$~MeV the $\pi N$ sigma term~\cite{lec}, 
$f_\pi = 92.4$~MeV the pion decay
constant, and $\chi(0,\vec{q}_\pi)$ the p-wave pionic
susceptibility. Since the virtual $\pi^+$ propagates in a triton-like medium,
we have $\rho_n = 2\rho_p$. In infinite nuclear matter with Fermi
momentum $p_F$, the p-wave pionic susceptibility $\chi(0,\vec{q}_\pi)$ can be
approximated by a constant for $|\vec{q}_\pi|\lesssim p_F$, and we will assume
that this is also the case here, although a local density approximation for
such a small nucleus may be questionable. 
With the two values for $\Sigma_\pi$ given
above, we immediately obtain $\chi = 0.31\pm 0.22$. On the other hand,
a standard 
calculation with particle-hole (ph) and $\Delta$-hole ($\Delta$h)
susceptibilities (see e.g. \cite{ErWe}) for infinite isospin-asymmetric
nuclear matter results already at small densities in much higher values for
$\chi$. For example, with $\rho_p + \rho_n = \frac{1}{3} \rho_0$ ($\rho_0 =
0.17$~fm$^{-3}$ being the saturation density), $\rho_n = 2\rho_p$
and the Migdal parameters
$g^\prime_{NN} = 0.8$ and
$g^\prime_{\Delta N} = g^\prime_{\Delta\Delta} = 0.6$, we find
$\chi\approx 0.8$ (Fig.~\ref{fig:pi}). 
This is principally due to the large contribution
of the ph 
Lindhard function which, at $\omega_\pi = 0$, is proportional to $p_F$ and
therefore does not change appreciably if one reduces the density within a 
reasonable range. One obvious improvement is the use of an energy gap 
in the ph-spectrum at the Fermi surface. It accounts in an average way for the 
low-lying excitation spectrum of a finite nucleus~\cite{OSTN}. Using a gap
of 8.5 MeV, appropriate for the continuum threshold of the triton, leads to
a reduction of $\chi$ but is still not able to describe the 
slope of $\Sigma_\pi$ inferred from the measurement (Fig.~\ref{fig:pi}). 
This indicates that  
the use of the bulk-matter Lindhard function is not appropriate for such 
a small nucleus and the kinematics probed in the experiment.
Therefore we do not attempt to calculate $\chi$ but rather use the above 
value $\chi = 0.31\pm 0.22$ from experiment. 
This allows an extrapolation of the self energy to
$\vec{q}_\pi = 0$
and to determine the mean density experienced by the virtual pion, with
the result $\rho = \rho_p+\rho_n =
\left(0.057^{\;+\;0.085}_{\;-\;0.057} \right)$~fm$^{-3} \approx 
\frac{1}{3}\rho_0$, albeit with a
large error. The self energy
corresponding to the best fit is displayed 
in Fig.~\ref{fig:pi}. 

For further physical interpretation of the measurement 
we use guidance from ChPT to infer the effective $\pi^+$ mass 
at the density probed in the present experiment. Given the above
mentioned uncertainties in the use of the local density 
approximation for the medium modification of the pion in very light
nuclei these results should be regarded as qualitative.  
The effective mass can be obtained from an extrapolation of the pion self
energy to the mass shell. Up to second order in $\omega_\pi$ and $m_\pi$, the
self energy of a charged pion in homogeneous, spin-saturated, but
isospin-asymmetric nuclear matter in the vicinity of $\omega_\pi\approx m_\pi$
and for $\vec{q}_\pi = 0$ is given by the expansion
\begin{eqnarray}
  \label{eq:pionself2}
  &&\Sigma_\pi^{(\pm)} (\omega_\pi, 0) =
    \left( -\frac{2\,(c_2+c_3) \omega_\pi^2}{f_\pi^{\,2}}
      -\frac{\sigma_N}{f_\pi^{\,2}} \right) \rho\\
  &&+\frac{3}{4\pi^2} \left( \frac{3\pi^2}{2} \right)^{1/3}
    \!\!\frac{\omega_\pi^2}{4 f_\pi^{\,4}} \,\rho^{4/3}
    \pm \frac{\omega_\pi} {2 f_\pi^{\,2}} \,(\rho_p - \rho_n)
    +\ldots\,,\nonumber
\end{eqnarray}
where the $+/-$ signs refer to the respective charge state of the pion (see
also~\cite{wei_nan}). The
low-energy constants (LEC's) $c_2$ and $c_3$ of the Chiral Lagrangian and the
$\pi N$ sigma term $\sigma_N$ characterize the $\pi N$ interaction and are
related to the $\pi N$ scattering lengths. We use $(c_2+c_3)\times
m_\pi^2=-26$~MeV~\cite{lec}, but one should remark here that third-order
corrections may change the LEC's somewhat~\cite{fetmei00}. The pion self
energy in Eq.~(\ref{eq:pionself2}) consists of two isoscalar parts
proportional to $\rho$ and $\rho^{4/3}$, respectively, and an isovector part
proportional to $(\rho_p-\rho_n)$. The latter is known as the
``Tomozawa-Weinberg term''~\cite{tomo}. Based on PCAC arguments, it reflects
the isovector dominance of the $\pi N$ interaction at $\omega_\pi = m_\pi$,
where the isoscalar scattering length as given by the first coefficient in
Eq.~(\ref{eq:pionself2}) vanishes at leading order. The
second isoscalar term proportional to $\rho^{4/3}$ is caused by
s-wave pion scattering from correlated nucleon pairs~\cite{eri-eri}. The sign
of the Tomozawa-Weinberg term depends on the isospin asymmetry of the nuclear
medium. In the present case of a virtual $\pi^+$ propagating in a triton-like
medium with $\rho_p - \rho_n = -\frac{1}{3}\, \rho$, the isovector term
becomes attractive.

\begin{figure}[t]
\centering\includegraphics[angle=90,width=\textwidth]{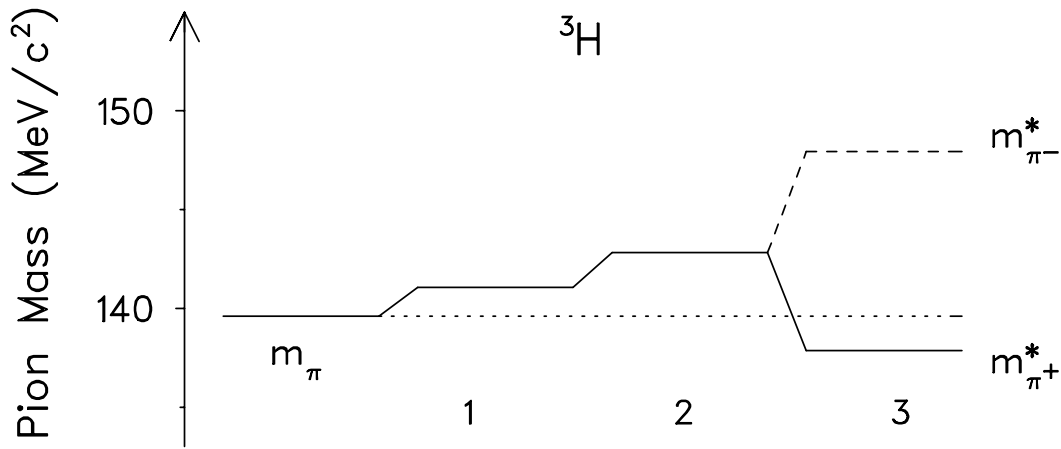}
  \caption{
The charged-pion mass shifts in the triton at the effective density
of the present experiment. 
Starting from the bare mass $m_\pi$ the first and the second term in 
Eq.~(\ref{eq:pionself1}) are repulsive
while the third term is
repulsive for $\pi^-$ and attractive for $\pi^+$ and causes the mass
splitting. 
}
\vspace{2ex}
  \label{fig:pionmassshift}
\end{figure}
The effective $\pi^+$ mass $m_{\pi^{+}}^{*}$
is deduced from the pole of the pion
propagator at $\vec{q}_\pi = 0$ which is determined by the solution of  
$\omega_\pi^2 - m_\pi^2 - 
\Sigma_\pi(\omega_\pi,0) = 0$ with the self energy as given by
Eq.~(\ref{eq:pionself2}).
Using $\rho= \left(0.057^{\;+\;0.085}_{\;-\;0.057}\right)$ 
fm$^{-3}$, one obtains a mass shift
$\Delta m_{\pi^+} = m_{\pi^+}^{*} - m_{\pi} = 
\left(-1.7^{\;+\;1.7}_{\;-\;2.1} \right)$ MeV/c$^2$ when the $\pi^+$
propagates in $^3$H.
It is interesting to compare the determined negative
mass shift $\Delta m_{\pi^+}$ 
with a positive mass shift $\Delta m_{\pi^-}$ derived from deeply
bound pionic states~\cite{yama,ita} 
in $^{207}$Pb and $^{205}$Pb with $N/Z \simeq 1.5$.
Itahashi {\sl et al.}~\cite{ita} 
have reported a strong repulsion of 23 to 27~MeV due to the local potential
$U_{\pi^-}(r)$ for a deeply
bound $\pi^-$ in the center of the neutron-rich $^{207}$Pb nucleus.
Evaluating Eq.~(\ref{eq:pionself2}) for this case with 
$\rho_p + \rho_n = \rho_0$
and $\rho_n / \rho_p = N/Z \simeq 1.5$
one calculates $U_{\pi^-}(0) = \Sigma^{(-)}_\pi(m_\pi,0)/(2 m_\pi)
\approx 18$ MeV. This is in good agreement with the findings of
Ref.~\cite{wei_nan}. Yet, there remains the problem of a 
``missing repulsion'' in the interpretation of the pionic atom data. 

Figure~\ref{fig:pionmassshift} shows the contributions to the
pion mass shift in $^3$H:  
The two isoscalar contributions to $\Sigma_\pi$
are both
repulsive and increase the pion mass. One thus notices 
from Eq.~(\ref{eq:pionself2}) that at $\omega_\pi
= m_\pi^* \neq m_\pi$ already the isoscalar contribution to the self
energy is sizeable. 
For a neutron-rich nucleus the isovector $\pi N$ interactions are
attractive (repulsive) for $\pi^+(\pi^-)$ giving rise to a splitting of
the mass shifts (contribution 3 in Fig.~\ref{fig:pionmassshift}). 
In $^3$H, the isoscalar and isovector terms are
compensating each other to a large extent, resulting in the very small
decrease of the $\pi^+$ mass. 

\subsection{Modification of the $\Delta$}
\label{sec:deltamod}
Most of the DWIA overestimate in the transverse channel
(cf. Fig.~\ref{fig:rosen}) is removed by a medium modification of the
$\Delta$ isobar.
The in-medium $\Delta$ propagator is written~\cite{kam_pi0he4} as
$[\sqrt{s} - M_\Delta + i\Gamma_\Delta/2 - \Sigma_{\Delta}]^{-1}$,
where one introduces 
a complex self-energy term $\Sigma_{\Delta}$ in the free $\Delta$ propagator.
Besides this explicit medium modification of the production amplitude also the
DWIA formalism for the pion-nucleus rescattering effectively accounts for a
$\Delta$ modification in the medium~\cite{kam93}.
The quantity  $\Sigma_{\Delta}$ has been deduced 
from an energy-dependent fit to
a large set of $\pi^0$ photoproduction data~\cite{kam_pi0he4,ram99} 
from $^4$He and also consistently describes recent photoproduction data from 
$^{12}$C, $^{40}$Ca and $^{208}$Pb~\cite{krusche}.
The fitting procedure reported in~\cite{kam_pi0he4} has been redone
with the unitary phase excluded from the propagator in accordance with
prescriptions often used in the $\Delta$-hole model~\cite{koch84}. 
The resulting $\Delta$ self energy exhibits a dependence on the photon energy.
Evaluated for the kinematics 1 and 2,
which correspond to the photon equivalent energies $k_{\gamma}^{eq} = 392$ and
376 MeV, respectively,
the real and imaginary parts are
$Re\, \Sigma_{\Delta} \approx 50$ and 39 MeV and 
$Im\, \Sigma_{\Delta} \approx -36$ and $-29$ MeV.
Although quite large values are obtained in view of the small density
$\rho \approx \frac{1}{3}\rho_{0}$, one should stress 
that the on-shell $\Delta$
self energy at resonance position is numerically considerably 
smaller~\cite{kohl,emi2001}.
As a result,
the agreement with the transverse cross section is significantly
improved, although the experimental values are still
overestimated by about 30\%.
The remaining discrepancy may be due to additional theoretical
uncertainties. For example, the Fermi motion of the nucleons is
effectively accounted for by a factorization
ansatz~\cite{kam_pi0he4}. An exact treatment might reduce the
prediction of the transverse cross section by about 10\%. A second
uncertainty of the order of 10\% concerns the knowledge of the
elementary $\pi^+$ production amplitude at $\theta_\pi =
0^\circ$. This kinematical region is not probed in
photoproduction but may be accessible in the future with appropriate
electroproduction data from the proton.
Attributing the entire $\Delta$ self energy to a mass shift $\Delta
M_{\Delta}$  and a width change $\Delta \Gamma_{\Delta}$, we deduce an 
increase  by 40 to 50 MeV and 60 to 70 MeV, respectively. 
These values seemingly differ from our earlier results~\cite{blo97}, 
where we have employed the  parameterization from Ref.~\cite{Car92} which 
did not include the $\Delta$-hole interaction,
giving $Re\,\Sigma_{\Delta} \approx -14$ MeV for a mean $^{3}$He density of
$\rho = 0.09$ fm$^{-3}$. On the other hand,
the self-energy term of the present work is an effective parameter
which incorporates the influence of the $\Delta$-spreading potential,
Pauli- and binding effects as well as the
$\Delta$-hole interaction including the
Lorentz-Lorenz correction. This finally leads to the positive sign.

\begin{figure}[t]
\centering\includegraphics[angle=90,width=\textwidth]{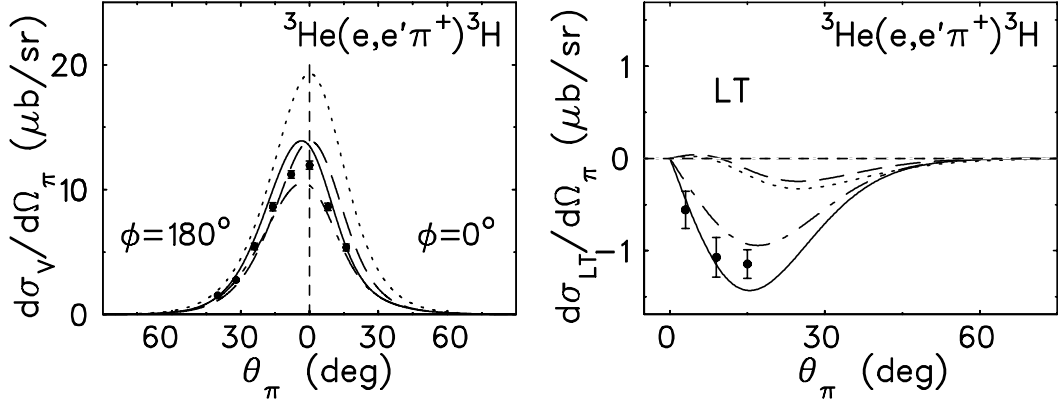}
  \caption{
L.h.s.:
The pion angular distribution measured at $Q^2 =
0.100$ (GeV/c)$^2$.  
R.h.s.:
The $LT$ term of the differential cross section.  
The labeling
is the same as in Fig.~\ref{fig:rosen}.
}
  \label{fig:angdistri}
\end{figure}

The effects of the medium modifications were also examined in the angular
distribution of the produced pions in kinematics 2. 
The data along with the model calculations 
are shown  in the l.h.s. of Fig.~\ref{fig:angdistri}. 
The asymmetry of the combined distribution
is due to a finite $LT$ interference term. 
From the azimuthal dependence on $\phi_\pi$ for three  polar angle
bins $\theta_\pi$
we extract the $LT$  interference 
term as a function of the pion emission angle $\theta_{\pi}$,
as shown in  the r.h.s. of
Fig.~\ref{fig:angdistri}  along with  the comparison  to
the model calculations.
It is obvious 
that only the full calculation, incorporating the medium modifications
in the pion and $\Delta$ propagators, is able to reproduce the angular
distributions. 

\section{Summary}
\label{sec:summary}
In summary, in a kinematically complete experiment, we have measured
the longitudinal and transverse cross section as well as the $LT$ 
interference term for the first time in the 
$^3$He(e,e'$\pi^+$)$^3$H reaction. The high sensitivity of the
electroproduction cross section shows clear 
evidence for self-energy corrections in both the pion and 
$\Delta$-isobar propagators and complements the large body of previous
results from pion-nucleus data. Using ChPT we have extrapolated the
pion self energy determined from the present experiment to the mass
shell to deduce the effective $\pi^+$ mass in $^3$H. Although
qualitative, the results appear to be consistent with the 
theoretical analysis of deeply bound pionic atoms
and the deduced effective $\pi^{-}$ mass~\cite{wei_nan}. 
In the transverse channel, the medium
modification of the $\Delta$ isobar is also evident and the 
self-energy modifications inferred from the present measurements
conform with $\pi^0$ photoproduction data over a wide mass range.

\section{Acknowledgement}
\label{sec:acknowledgement}
This work has been supported by the Deutsche Forschungsgemeinschaft
(SFB 443 and RI 242/15-2).
We are very grateful to P.~Kienle who made a remark to one of us
(A.R.) on the pion mass which finally led to the results presented
here, and to N. Kaiser and W.~Weise for stimulating discussions.



\end{document}